\documentclass[journal]{IEEEtran}
\ifCLASSINFOpdf
\else
\fi
\usepackage{amsmath}
\usepackage{amsfonts}
\usepackage{graphicx}
\usepackage{graphicx}
\usepackage{xcolor}
\usepackage{cite}
\usepackage{tasks}
\usepackage[bottom]{footmisc}
\usepackage{epstopdf}
\usepackage{subcaption}
\usepackage[left=0.55in,right=0.55in,top=0.6in,bottom=0.6in, textheight=18in]{geometry}

\makeatletter
\def\BState{\State\hskip-\ALG@thistlm}
\makeatother

\begin{document}
\raggedbottom

\title{Security of 5G-V2X: Technologies, Standardization and Research Directions}
% with support for Augmented Reality

\author{Vishal Sharma, Ilsun You, Nadra Guizani
\thanks{V. Sharma and I. You (Corresponding Author) are with the Department of Information Security Engineering, Soonchunhyang University, The Republic of Korea, Email: vishal\_sharma2012@hotmail.com, ilsunu@gmail.com. N. Guizani is with the Washington State University, USA, Email: nadraguizani07@gmail.com. This work is supported by the Soonchunhyang University Research Fund.}% <-this % stops a space
}

% space would be appended to the last name and could cause every name on that
% instance, "\textbf{A} \textbf{B}" will typeset as "AB" not "AB". To get

\markboth{}%
{}
% The only time the second header will appear is for the odd-numbered pages
% after the title page when using the two side option.



\maketitle

\begin{abstract}
Cellular-Vehicle to Everything (C-V2X) aims at resolving issues pertaining to the traditional usability of Vehicle to Infrastructure (V2I) and Vehicle to Vehicle (V2V) networking. Specifically, C-V2X lowers the number of entities involved in vehicular communications and allows the inclusion of cellular-security solutions to be applied to V2X. For this, the evolvement of LTE-V2X is revolutionary, but it fails to handle the demands of high throughput, ultra-high reliability, and ultra-low latency alongside its security mechanisms. To counter this, 5G-V2X is considered as an integral solution, which not only resolves the issues related to LTE-V2X but also provides a function-based network setup. Several reports have been given for the security of 5G, but none of them primarily focuses on the security of 5G-V2X. This article provides a detailed overview of 5G-V2X with a security-based comparison to LTE-V2X. A novel Security Reflex Function (SRF)-based architecture is proposed and several research challenges are presented related to the security of 5G-V2X. Furthermore, the article lays out requirements of Ultra-Dense and Ultra-Secure (UD-US) transmissions necessary for 5G-V2X.
\end{abstract}

% Note that keywords are not normally used for peer review papers.

% \fi
%
% For peer review papers, this IEEEtran command inserts a page break and
\IEEEpeerreviewmaketitle

\section{C-V2X: Overview}%5G-Enabled Automated Cars
The current era of vehicular travel is witnessing a dynamic shift from individually driven vehicles to network controlled vehicles. This new infrastructure is studied under the name of Vehicle-to-Everything (V2X), which aims at controlling vehicular communications for specific operations where a vehicle is able to communicate with any of the other entities available on the network. Growing from Vehicle-to-Infrastructure (V2I) and Vehicle-to-Vehicle (V2V), V2X broadens the domain of its applicability while leveraging on a different range of technologies, such as Dedicated Short-Range Communications (DSRC), Wireless Access in Vehicular Environment (WAVE), Cellular-V2X (C-V2X) that includes Long-Term Evolution V2X (LTE-V2X), 5G Infrastructure Public-Private Partnership (5GPPP-V2X), automated-Ethernet (onboard communications), and Wireless Local Area Network V2X (WLAN-V2X)~\cite{8300313}~\cite{wang2019platoon}~\cite{earls2017wlan}. C-V2X is also seen as a base for implementing technologies like Low Power Wide Area Network (LPWAN), IPv6-Low-Power Wireless Personal Area Network (6LoWPAN) and Long Range Wide Area Network (LoRaWAN) where conservation of energy is the primary motive of the deployed technology.

It has been predominantly established by earlier studies that C-V2X is a better alternative to any of the existing technologies based on the performance and deployment strategies. However, factors like coverage, mobility management, Total Cost of Ownership (TCO), reliability, latency, security, and scalability are yet to be evaluated based on the existing infrastructure~\cite{ma2016v2x}\cite{8447062}. There has been a huge rush towards the establishment of LTE-V2X models while studying the capability of LTE in terms of performance and security. The primary motive for utilizing existing strategies is for their communication range. To provide security for these models, additional functional layers are added. This not only increases the cost of ownership but also decreases the compliance of autonomous vehicles~\cite{5Gamericas}~\cite{3GPPTS}.

%Moreover, there is a gap in the studies which need to follow the realistic urban model for implementation and test applications based on upcoming standards for C-V2X.
In short, C-V2X aims at bridging the gap between the vehicular and cellular communication industry by supporting a large range of Information and Communications Technology (ICT) applications. All the major technologies targeting C-V2X can be observed for the following categories:
\subsection{Multi-Vendor Services Support (MVSS)}
V2X depends on the convergence of a large number of cellular-applications, which are being provided by multiple vendors. In general, C-V2X is considered to be an operational property of a single organization (Original Equipment Manufacturer (OEM)), which uses cellular facilities to control the transmissions in the network. However, with a variety of cellular-applications, it is liable that a single vehicle will be supported by multiple vendors. Thus, Multi-Vendor Services Support (MVSS) becomes one of the crucial principles to be followed in C-V2X. Layouts through slices, edge-formations, fog-infrastructure, Software Defined Networking (SDN), and Network Function Virtualization (NFV) can be the principal technologies for MVSS~\cite{campolo20175g}. %Zero-delay policing, smooth and fast handovers, mutual authentication and service layoffs are the major factors to control. Interference management, power allocation, Line of Sight (LoS), and Non-Line of Sight (NLoS) are other issues to be handled in case of multiple services to the same vehicle.
\subsection{Autonomous Algorithm Safety (AAS)}
Algorithms are the key behind the successful operations of autonomous vehicles in C-V2X. Majority of these algorithms rely on the formation of a secure channel between the vehicles (V) and all other applications in the network (X). Vulnerability in the algorithms can lead to several types of cyber attacks on C-V2X. The threat level increases as vehicles in the network operate from full-assistance to no-assistance (fully-autonomous). As discussed in~\cite{8300313} by the Society of Automotive Engineers (SAE), the AAS depends on the mode of operations and deployment scenarios of vehicles. Specifically, in C-V2X, channel security, session management, security-patches, key management, access control, and camouflage-detection are the key perspectives to look forward to for AAS. Policing, resource management, and risk mitigation are other issues to be tackled for AAS in C-V2X.
%In addition, the control is exhibited through the underlying communication technology, but the safety is based on the similar properties as that of AAS.
\subsection{Network Control and Safety (NCS)}
From the C-V2X point of view, it is required to study network control and safety as a single component, as their tradeoff shows a considerable impact on the implementation as well as the security of the network. Attaining MVSS and AAS helps to efficiently control the operations in C-V2X. The detection of anomalies, attack-mitigation, and prevention against zero-day vulnerabilities are other metrics that need to be efficiently handled. NCS also accounts for the management of vulnerable activities, misbehavior detection and session security in C-V2X.
\section{Use-Cases for Secure C-V2X}
C-V2X aims at facilitating on-the-go network, which primarily matches similar capabilities of a stand-alone cellular network. Several studies are available that highlight the practical aspects and application-based use-cases of C-V2X, however, in order to complement the existing findings and studies, some use-cases from the C-V2X security perspective are listed below:
\begin{itemize}
  \item Autonomous Car Security: Autonomous cars use real-time data and instructions from different sensors connected to the cellular network. The guidance maps for real-time coordination can be accessed through the C-V2X communications. The security features of C-V2X help to prevent any impersonation and replay attack which may misguide the vehicle and lead to interruptions and accidents. The security considerations and applying several key-based mechanisms can help to provide strong encryption for transmissions involving guidance data to autonomous cars.
  \item Driver Authentications: In assisted cars, secure operations of C-V2X can help verify the drivers through third-party authentications. The medical conditions of the driver can also be verified through attached sensors and several light-weight authentications can help to quantify access control to the legitimate driver.
  \item Vehicle-Health Monitoring: The vehicle's health can be monitored through C-V2X, which sends instructions in real-time to car software maintainers for every machine issue. During wrong- configurations, there are high possibilities for an intruder to gain access to the components of a vehicle which may be further exploited to gain access to the entire network. Such situations can be encountered through the formation of a secure communication channel in C-V2X.
  \item Secure Public Safety Communications (PSCs): C-V2X is expected to play a pivotal role in PSCs by allowing vehicles to communicate the shortest path to other devices in cases such as: real-time delivering of food, medicine, and another kind of services that are time-sensitive. Moreover, the security features of C-V2X can help extend its applications to military and civilians expeditions. The systematic and secure coordination can help to attain high reliability and low latency for the devices involved in C-V2X setup.
  \item Inter and Intra Vehicular Security: Trust and privacy are major concerns in the case of inter- and intra-vehicular communication~\cite{zhang2017security}. Inter-vehicular communications refer to the C-V2X setup that is comprised of vehicles from different vendors. In such a scenario, security becomes a dominant factor, and it is expected to use pseudonyms or proxies for preventing inter-network eavesdropping. Intra-vehicular communication refers to the onboard operations of a vehicle involved in C-V2X. In this mode, security technologies are required to protect the customer's private information and vehicle systems from hackers. The attacks in any of these modes pose a considerable effect on the trustworthiness of the network as well as creates a huge impact on ownership and quality of experience.
  \item Secure Mobility management and Service Layoffs: Mobility and service layoffs are the crucial aspects of C-V2X. There are several solutions available that focus on both fast and secure service layoffs and handover management~\cite{sharma2018secure}. However, with proprietary network formations, the security factors become dominant and should be resolved through mechanisms applicable to C-V2X, especially leveraging on LTE and 5G technologies.
  \item S-B2MP (Secure Base to Multi-Peer Networks): One of the interactively keen examples of C-V2X is B2MP, in which a vehicle serves as the Base Station to multiple peers. However, the presence of an attacker on-board may expose the key metrics of the network, which can be used to launch several exploitations leading to a huge impact on the overall formation of C-V2X. Thus, S-B2MP is another security-oriented use case of these networks.
  \item Secured Named Data Networking (NDN): NDN is a basic network communication mode that supports secure data directly at the network layer by making every data packet verifiable. NDN uses ad hoc and broadcast-style communications and is independent of communication technologies. Therefore, it can be used with C-V2X to enhance its feature based on secure media-independent formations.
  \item Traffic Management and Anomaly Detections: The traffic management includes issues related to speed management, traffic information, routing information, cooperative navigation, etc~\cite{hobert2015enhancements}.  Moreover, driver-behavior, vehicular-anomalies, and network intruders are other factors affecting the core functionalities of the vehicular system. Sufficiently secure mechanisms can help to resolve these issues and identify potential anomalies prior to their attack. %However, detection of such activities is tedious and it is required that existing filtering algorithms must be experimentally verified before their application to C-V2X.
\end{itemize}

\section{C-V2X Security Architecture and Trends}
The traditional technologies for V2X, evolving V2I, and V2V, like DSRC and 802.11p, require a large number of entities for connecting vehicles and supporting their transmissions to OEMs. However, with the evolution of C-V2X under 3GPP, the existing cellular infrastructure can be used for supporting vehicular communications. With the advent of LTE technology, the V2X is highlighted for its vast range of applications, and with a shift of LTE towards 5G, the upcoming trends are focusing on utilizing both these architectures to provide security services to the involved entities\cite{molina2017lte}~\cite{5Gamericas}~\cite{3GPPTS}. This section discusses base architectures defined for LTE-V2X and 5G-V2X along with their security concerns and applicability.
\subsection{LTE-V2X}
\begin{figure}
  \centering
  \includegraphics[width=270px]{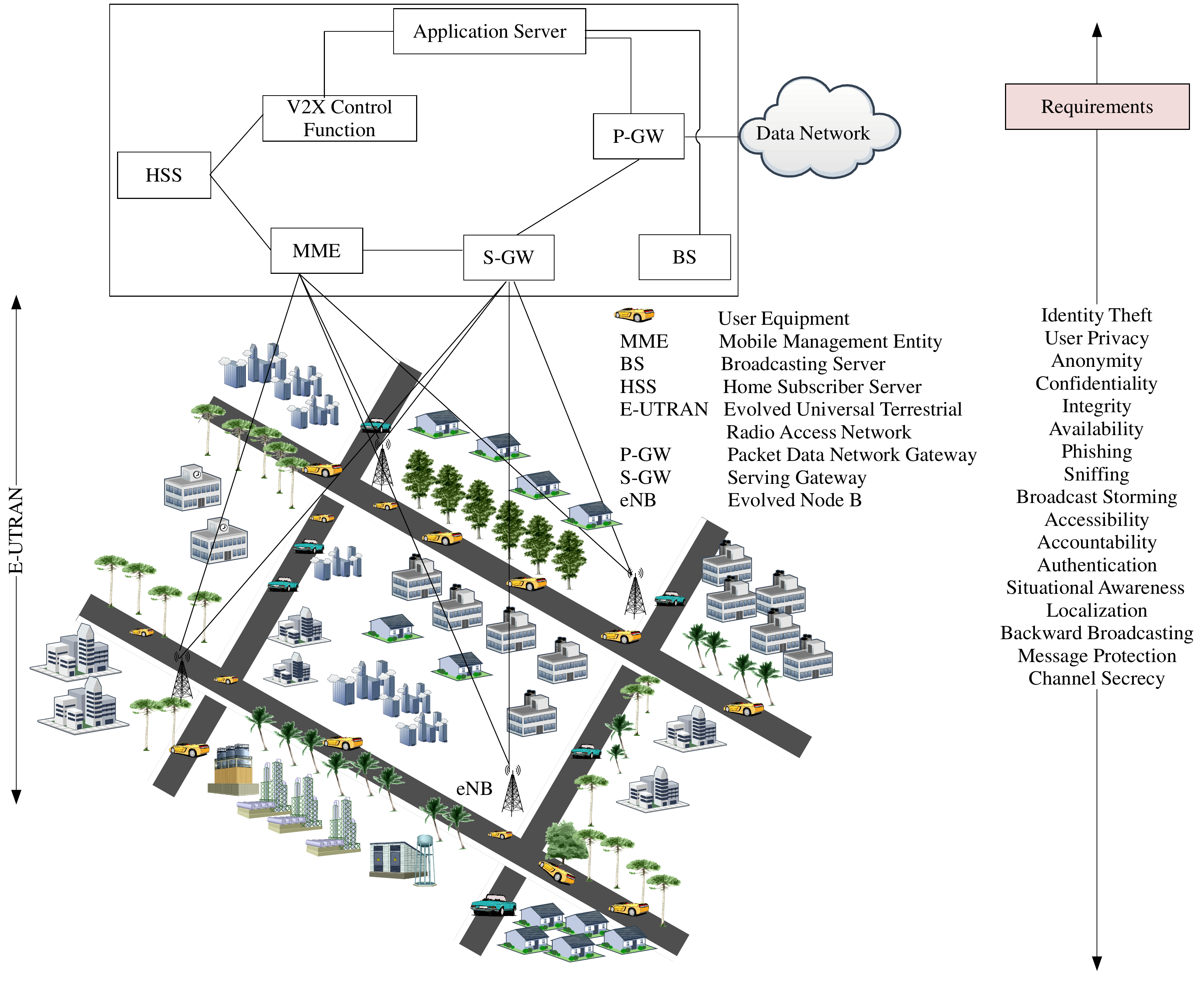}
  \caption{An illustration of LTE-V2X architecture and security requirements~\cite{5Gamericas}.}\label{fig1}
\end{figure}

LTE-V2X leverages services from eNB, and Mobility Management Entity (MME), which accounts for providing various control functions for V2X, as shown in Fig.~\ref{fig1}. The standard LTE architecture is comprised of Packet Data Network Gateway (P-GW), User Equipment (UE), Serving Gateway (S-GW), Home Subscriber Server (HSS), and Broadcasting Server (BS). All of which operate as components of the Evolved Universal Terrestrial Radio Access Network (E-UTRAN)\cite{8454322}. The traffic to the Data Network (DN) is facilitated through P-GW and measurement report based transmissions are used for supporting vehicular communications. Although efficient, there are a series of issues with this architecture as it is unable to provide strong mechanisms for vehicle authentication, credential management, privacy and anonymity of involved entities~\cite{bian2018toward}~\cite{zhang2017security}. Moreover, LTE-V2X does not comply with the upcoming requirements of ultra-low latency and ultra-high reliability~\cite{5Gamericas}. In addition, there is limited support for positioning and trajectory-based solutions for V2X. As depicted in Fig.~\ref{fig1}, the security requirements are tedious to resolve based on the component architecture of LTE. Thus, a paradigm shift is required from LTE to 5G for supporting edge computing, which is an integral aspect of V2X services.
\subsection{5G-V2X}
In contrast to LTE, 5G-V2X is a function based architecture which primarily focuses on providing service-based accessibility to the involved entities. The key advantages of 5G-V2X are service-based policing for applications, low-latency, high-reliability and functional support for V2X, which can be operated in Non-Standalone 5G (NS-5G) or Standalone-5G (S-5G) mode depending on the deployment changes to the initial architecture~\cite{3GPPTS}\cite{chen2017vehicle}. NS-5G-V2X is dependent on the underlying LTE-deployment to facilitate the requirements laid by 5G communications. The scope of enhancement to security is limited as this requires the exact identification of the 5G functions, which will match the components defined in LTE~\cite{5Gamericas}. However, with S-5G, the scope widens, but it also requires work from the ground level while managing communication back to the core.

\begin{figure}[!hb]
  \centering
  \includegraphics[width=260px]{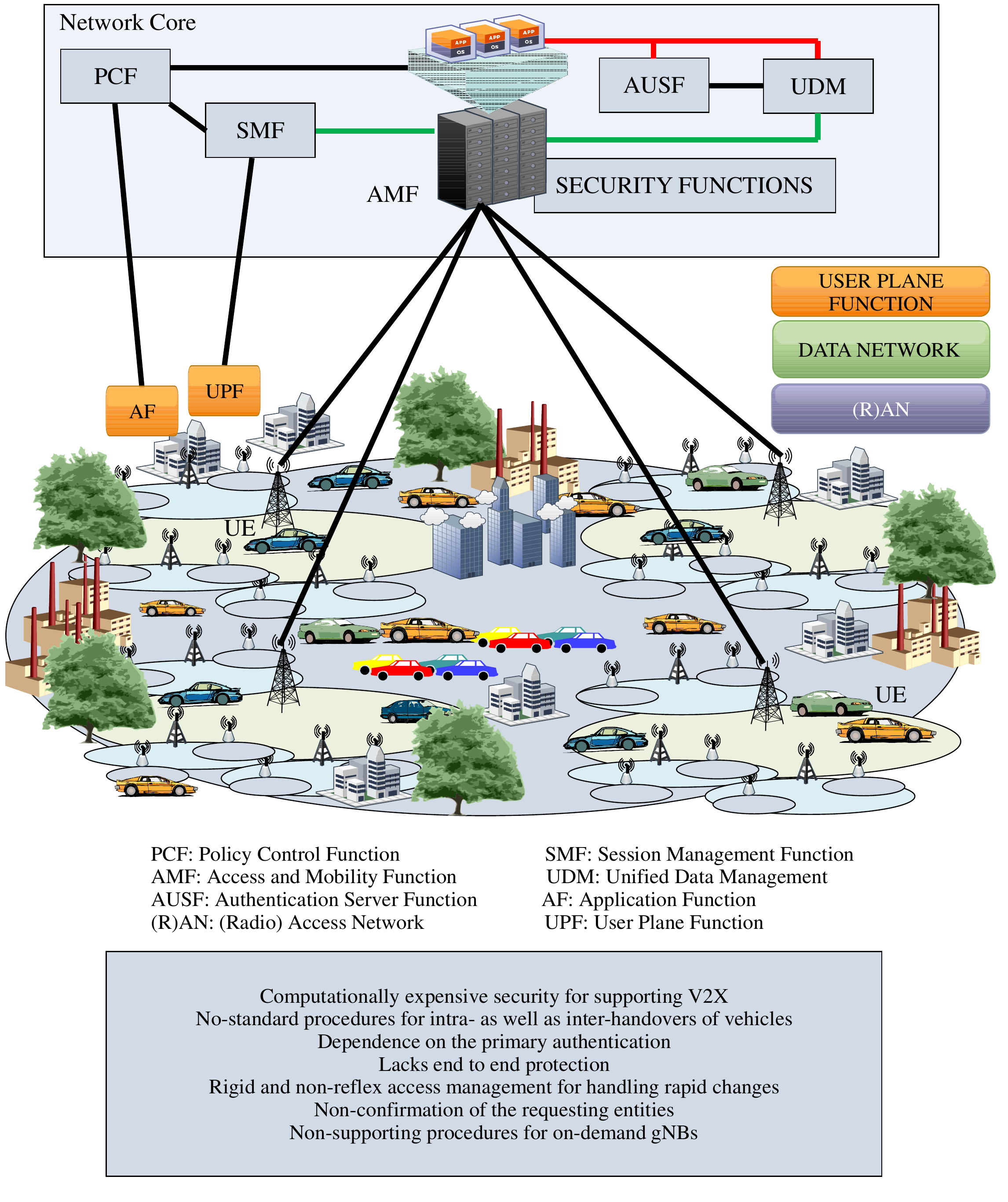}
  \caption{An illustration of exemplary scenario for 5G-V2X architecture based on 3GPP TS 23.501.}\label{fig2}
\end{figure}

The core functions of 5G-V2X setup include; Policy Control Function (PCF), Access and Mobility Function (AMF), Authentication Server Function (AUSF), Session Management Function (SMF), Application Function (AF), Unified Data Management (UDM), and User Plane Function (UPF), as shown in Fig.~\ref{fig2}. The security features are provided through specified security functions, namely, Authentication Credential Repository and Processing Function (ARPF), and Security Anchor Function (SEAF) both of which are collocated with the AUSF\cite{chen2017vehicle}. The details on each of them can be followed from the technical specification by 3GPP on the security of 5G networks~\cite{3GPPTS}. However, there are no concurrent studies which discuss the security from V2X perspectives. This article provides an initial screening of such requirements as discussed in the next section.

\section{5G-V2X: Security Attacks, threats, and requirements}
The majority of the shortcomings of DSRC, 802.11p, and LTE-V2X are meant to be handled through the efficient function handlers in 5G-V2X. The search for security solutions and possible remedies against known and unknown threats depend on the deployment strategies of 5G-V2X. If V2X is enabled with NS-5G, the attacks possible on LTE-V2X holds true and can exploit the services in 5G-V2X. However, with S-5G, the attack window decreases and the protection against threats can be increased while maintaining ultra-low latency and ultra-high reliability amongst the entities. Network planning and deployment play a key role in deciding the security of 5G-V2X. The placement of functions and control, and decisions on policing implicate possible exploitation of vulnerabilities. In addition, the exposure of keys and the use of an insecure channel of communication are other points of attack in 5G-V2X. Moreover, V2X forms the edge component of 5G, which may or may not have a secure channel. Thus, the possibilities of attacks increase when the devices undergo major mobility transitions. The cell coverage is another issue that can be evaluated by the eavesdropper to launch any potential attacks on the vehicles.
\subsection{General Issues}
This section provides detail for several key attacks and threats when deploying services through 5G-V2X.
\begin{itemize}
  \item The main reason for a possible attack in 5G-V2X is the irregular placement of gNB, which is the counterpart of eNB and MME of LTE-V2X. The primary impact can be caused by the authentication and authorization of vehicles. In the semi-autonomous mode, a certificate-based security, provided through email or semi-autonomous mode, is used assuming the network operates on a secure-line up between RSU and OEM. However, with major autonomy, the certificate-based solution may hinder the smooth transit between gNBs.
  \item The presence of a malicious node may exploit the vulnerability in OBU and gain access to the network (zero-day attacks). Thus, it becomes the responsibility of the network entity to prevent such attacks. Static information and weak hash functions may lead to certificate forgery. Which prevents the capture of secure elements of a vehicle is an ultimate requirement.
  \item For cellular-assisted autonomous driving, it is desirable to prevent any known and chosen plain/ciphertext attacks. Such attacks are possible as major sensor information is shared without encryption. Backward broadcasting and signal storming are the other issues related to the security aspect of 5G-V2X.
  \item Message security is another factor for securing transmissions in 5G-V2X. The content in these networks should be secured through secret keys. With the existing security modules, the keys are generated by following a hierarchical pattern. Fresh keys need to be maintained, and synchronized patterns must be used to prevent any replay attack or De-synchronous attacks.
  \item Irrespective of the network planning and layouts, side-channel attacks are tedious to detect and can exploit the entire network by merely affecting the vehicle or gNB in the 5G setup. In addition to these, service-based attacks are expected to prevail in 5G-V2X unlike DSRC or LTE-V2X as all the content in the 5G is expected to be classified into several services. Thus, service-based attack prevention and threat detection are key issues to focus on while securing the functionalities in 5G-V2X.
\end{itemize}
It is worth noting that the security in 5G-V2X not only depends on the security functions but also on the location of certain regular entities/functions, which involve gNB, SMF, AMF, and UPF. Control over any of these exploits the entire network. Thus, it becomes inevitably important to secure the passes between these entities while leveraging the services of security functions. However, the positioning of servers providing security functions must be carefully selected. A security anchor function near to a user may lead to several client-side attacks while placing at the core increases the latency and weakens the links between the AMF, SMF, and UPF.
%5G-V2X aims to address the issues of low-latency as well as high reliability while forming security solutions for a large set of applications.
\begin{figure*}
  \centering
  \includegraphics[width=380px]{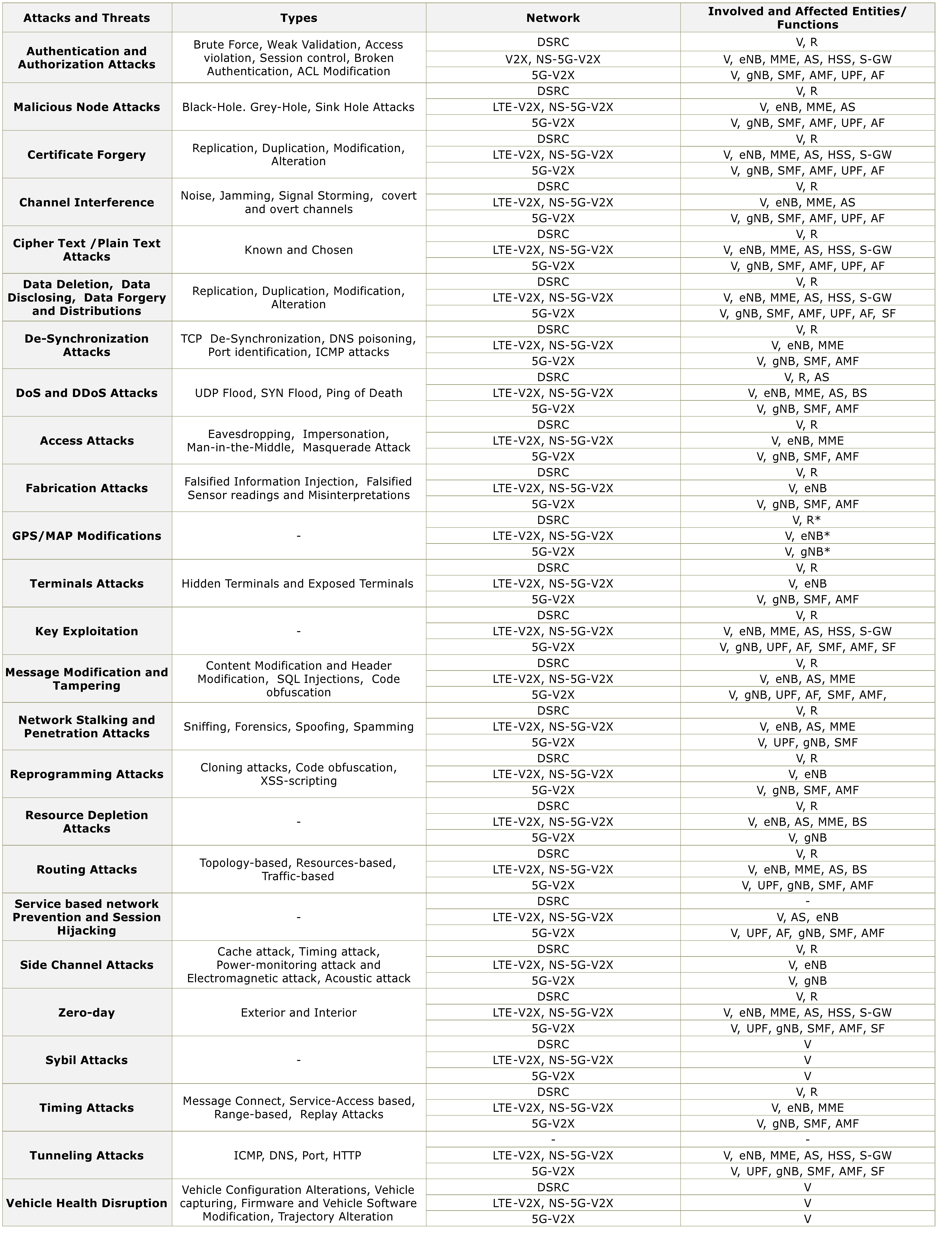}
  \caption{A detailed list of attacks and threats with a focus on the involved and affected entities/functions of C-V2X. (Entity List: - V: Vehicles (on-board units (OBUs))/ User Equipment, R: Road Side Units, AS: Application Server, SF: Security Functions)}\label{fig3}
\end{figure*}

\begin{figure*}
  \centering
  \includegraphics[width=340px]{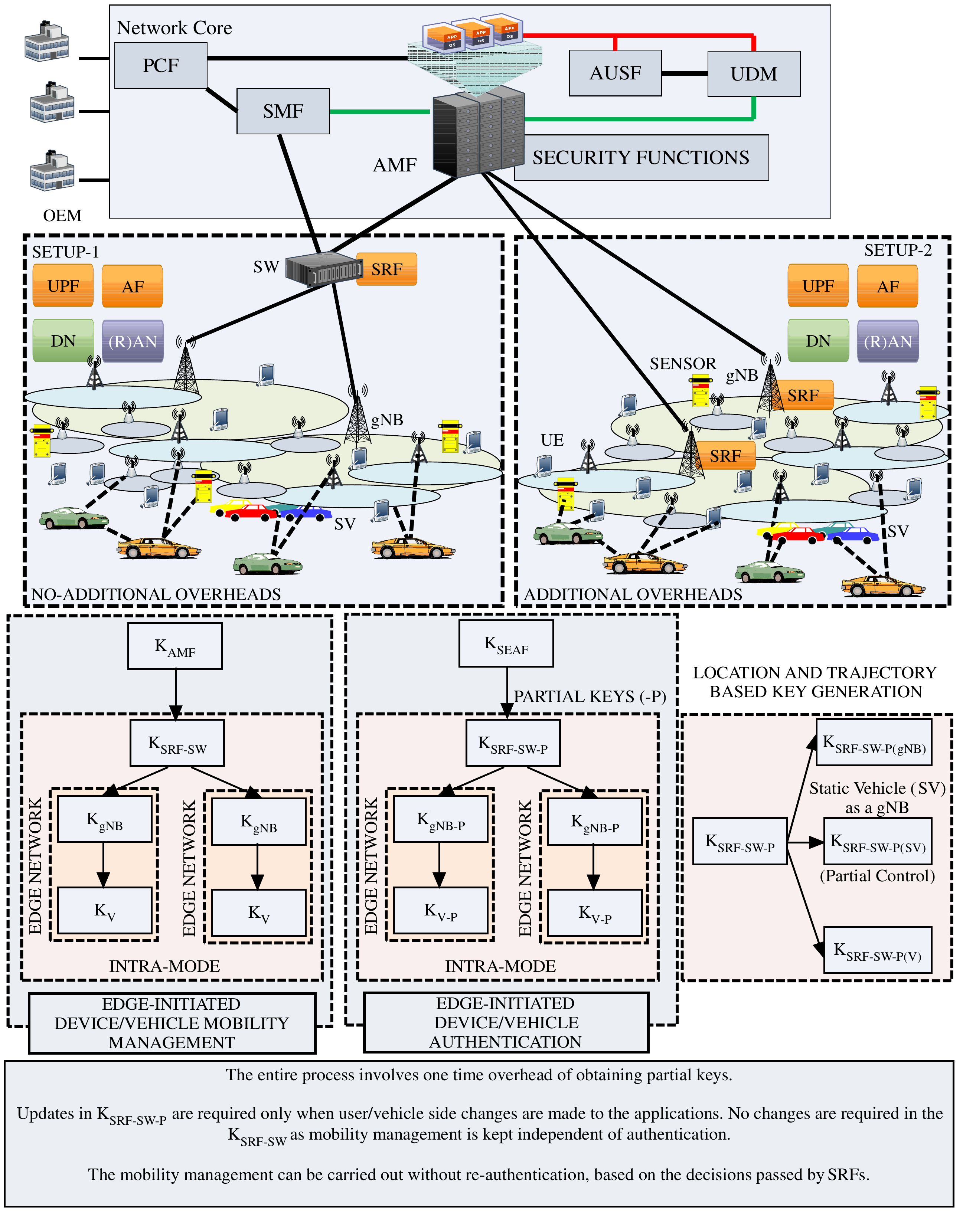}
  \caption{An illustration of the conceptualized 5G-V2X security architecture for intra- and inter-mode of operations for vehicles using Security Reflex Function (SRF).}\label{fig4}
\end{figure*}
\subsection{5G-V2X Specific Security Issues}
There is a scarcity of studies and no-concurrent solutions available which predominantly depicts the security aspects of 5G-V2X. The ones for NS-5G-V2X only focus on the existing issues limited to the infrastructure support of LTE. In the recent releases of TS series by 3GPPP~\cite{3GPPTS}\footnote{3GPP TS 33.501 V0.7.1 (2018-01)}, security is defined for 5G-V2X in the Access Stratum and Non-Access Stratum mode. The primary security is defined using 5G-AKA or EAP-AKA' through a hierarchical key distribution. The security is governed through secure key exchanges and by assuming different strategies for each of the involved entities. Although this report provides a detailed possible layout of security for 5G, it is yet to be considered for V2X because of differences in the dynamics and mode of operations of a vehicle from a regular UE.

V to X authentication and securing the credentials are the key issues to be considered for 5G-V2X~\cite{5Gamericas}. Moreover, network layout, planning, and handover are yet to have proper solutions for both intra- and inter-modes. The deployment of 5G-functions near to edge or core also needs further research from a 5G's perspective. Although, C-V2X (LTE and 5G) decreases the number of RSU required in the existing technology focused by vendors (from OEM to Vehicle connectivity), yet there are still issues pertaining to universal availability, interim-management of slices, and access management. The requirement of dynamic RSUs, as stated in ~\cite{5Gamericas}, can be attained through stationed vehicles, but there is no architecture to grasp this facility.

In addition to the above discussions, the 5G security reports depend on the expensive backward operations which become complex when applied to V2X solutions. Moreover, the use of a Certificate Revocation List (CRL) for initial authentication can only be accounted for a dense RSU network, and it involves a high dependency on a centralized authority, which is a problem when looking at a global deployment of unified V2X technology. The available information in TS reports~\cite{3GPPTS} resolves the perfect forward security for 5G UEs, but there is a gap in the use of this technology for V2X. Although, 5G architecture aims to protect keys used in the next phase, capturing of the vehicle or signature replication can lead to the violation of forwarding secrecy. Thus, attaining perfect forward secrecy is a crucial aspect for V2X because of the physical threats to the credentials of vehicles.

Another issue to be taken care of is the extensive dependence on the primary authentication and assumption of security assurance schemes. With the involvement of long-term secret keys, it is yet to be decided whether these will be generated through the 5G-core or the 5G functional units deployed in the periphery of OEMs. The protection of long-term keys depends on the deployment range and positioning of 5G-security functions for V2X. SEAF must be placed in the deep network leading to nearly impossible physical attacks, but this also raises the concerns as SMF and AMF, in this case, have to be placed near gNB or vehicle for facilitated transitions~\cite{3GPPTS}. The current versions do not provide any discussion on home network security of V2X and there are limited discussions on using public-key operations when the vehicle is operating in its home network. Also, attaining end to end protection by preventing Sybil attacks leading to effects on confidentiality and integrity is a must while deploying 5G-V2X solutions. Finally, the confirmations of requesting entities and identification of vehicles need to be decided on measuring both performance and security. For a clearer understanding, the impact of several attacks and threats with a difference in the use of technology is presented in Fig.~\ref{fig3}.
\section{5G-V2X: Conceptualized Architecture}
The technology solutions for C-V2X at the moment support more of V2V and V2I than V2X. The existing conceptualized views leverage 5G security modules for securing V2X communications. However, as discussed in the earlier section, the security for the majority of the components is done through computationally expensive operations and any sort of attack can cause severe damage. To resolve such issues and to further enhance the performance, a conceptualized architecture is proposed on the backbone of the architecture given by 3GPP~\cite{3GPPTS}.
%Thus, it is desired to re-model the existing architecture while facing primarily on the communication of vehicle with everything.
\subsection{Architectural Enhancements}
The proposed architecture discusses the security inclusions through edge computing where users, vehicles and several sensors/devices are treated as a part of everything and strategies are provided for both intra- as well as inter-handover of vehicles. The proposed conceptualized architecture uses a new function ``Security Reflex Function (SRF)", as shown in Fig.~\ref{fig4}, to support rapid changes in the network as well as to define policies for access management. Moreover, SRF accounts for attaining the feature of Ultra-Dense and Ultra-Secure (UD-US) mobility management, which is needed as it is expected that a huge number of cellular-supported vehicles will be roaming on roads demanding all time connectivity. It is desired to understand the features and operational strategy of SRF before following its role in 5G-V2X. The details are:
\begin{itemize}
  \item Edge-based authenticator: SRF provides edge-initiated authentication for the entities involved in 5G-V2X. It reduces the burden of the core by covering user-side roles of SMF and AMF.
  \item Partial-key allocations: SRF uses partial key allocations by deriving several keys from the keys obtained from AMF and SEAF. It uses device-based specific keys for managing V2X connectivity. It sits on top of gNB and can operate in a dual-mode with the specified gNB.
  \item Supports on-demand gNB: SRF allows strategic control over the network by including static vehicles as user-side gNBs, termed as gNB'. This also helps to support parking-based networks as well as Emergency Communication Vehicles (ECV).
  \item Allows splitting and slice management: SRF supports the core principle of slice management and helps to maintain the vehicle as well as slice anonymity by deriving several short-term keys depending on the mode of operations (intra or inter).
  \item User-side secondary authentication: SRF allows user-side authentication when the static vehicles are used as access points. Moreover, it allows secondary authentication for specifying route optimization by reducing the number of intermediate hops while maintaining the end to end security.
  \item Multi-radio facilities: With vehicles in proximity to everything, it is desired that multiple radio facilities must be supported by the 5G security functions. However, it is an expensive operation to include such facilities on all devices. Thus, SRF act as a common function, which allows radio-translations to support the security of vehicles having communication in different modes.
\end{itemize}
\subsection{Workflow and Key Generations}
The workflow and key generations in the conceptualized architecture can be orchestrated through specific frameworks or by simply dividing the existing keys. In the derived setup, partial keys are used, which can be treated similarly to the secondary authentication where SEAF is used to derive several SEAF'. However, SEAF' does not account for rapid changes, nor does it provides any support for edge-based V2X security. Additional overheads are also accounted for because of re-verification between the SEAF and SEAF'. In the proposed architecture, two different setups can be used to deploy SRFs.

In the first setup, the SRF can be fixed using switch (SW)-hub (gNB) architecture. The SRF then becomes the interface between the gNB and the core security functions and $K_{AMF}$ or $K_{SEAF}$ is used to derive several $K_{SRF-SW}$ and $K_{SRF-SW-P}$ keys, which further generate the $K_{gNB}$ and $K_{V}$ for the terminals and vehicles in its periphery. The derived keys are particularly applicable either for mobility management or authentication, as shown in Fig.~\ref{fig4}. This is the simplest form and it allows easier intra-handovers without additional overheads on gNB. However, it involves additional switches to be placed as a control center for several hubs or gNBs.

In the second setup, SRF and gNB are collocated, which adds to the overheads of operations on a single terminal. However, several security-passes and inclusion of additional switches as well as modifications to the core architecture can be avoided in this case. As an abstracted view, SRF may look like a derivative of existing architecture, but it provides a specialized location and trajectory-based key generations, which adds up to the efficiency and service-based security requirements of 5G-V2X. Based on the location as well as the availability of Static Vehicles (SV), SRF keys are used to derive additional keys, $K_{SRF-SW (SV)}$ and $K_{SRF-SW (V)}$(Fig.~\ref{fig4}), which enable 5G-V2X architecture to use SV as one of the gNBs. This widens the coverage and can be considered as one of the core solutions for PSCs through 5G-V2X\footnote{The proposed scheme divides the mobility and authentication procedures allowing the network to respond quickly to the rapid changes.}.

\begin{figure}[!hb]
  \centering
  \includegraphics[width=270px]{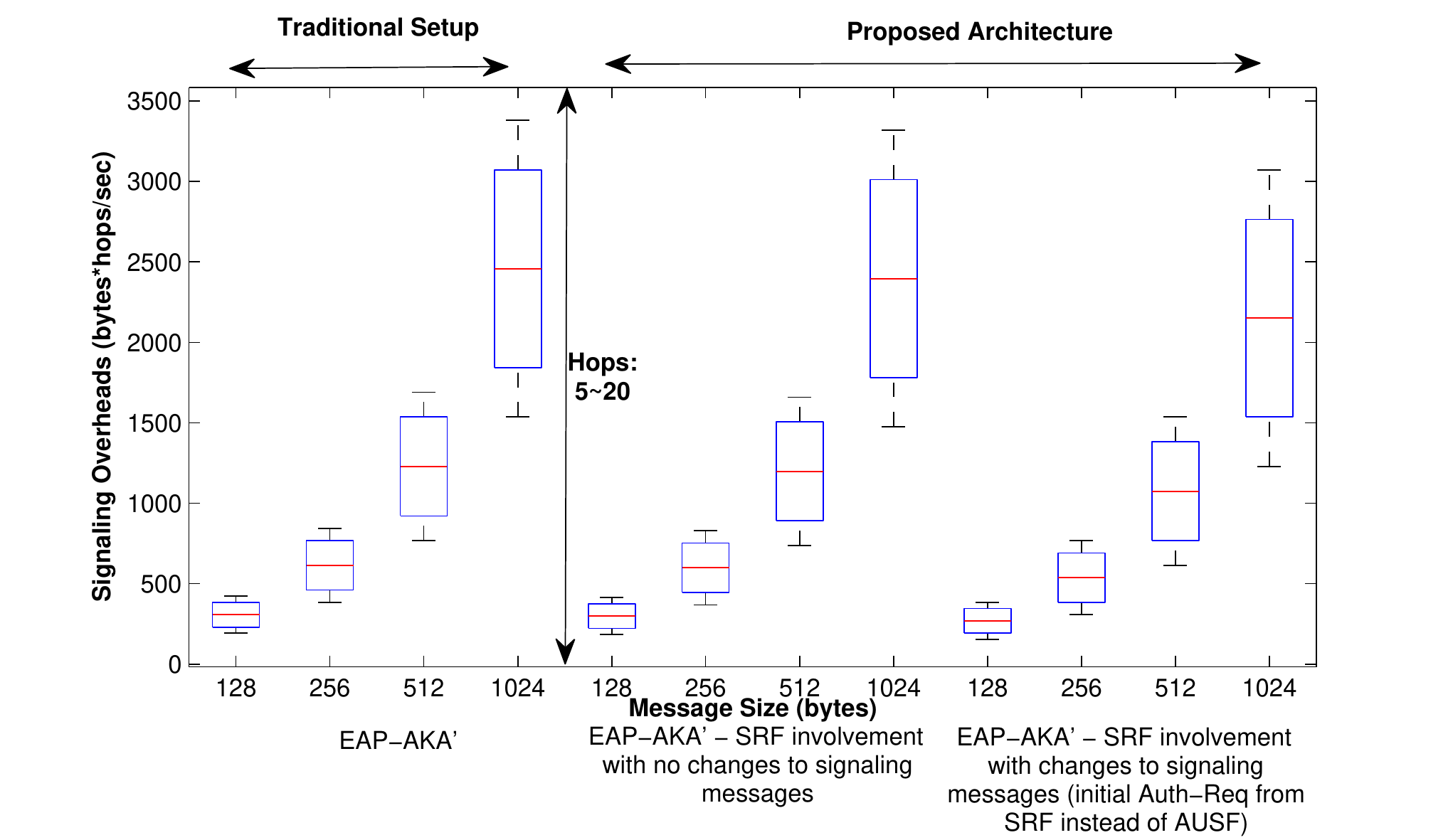}
  \caption{A graphical comparison for signaling overheads of vehicle's mobility through EAP-AKA' in the traditional and the proposed SRF-based 5G-V2X.}\label{fig6}
\end{figure}

\begin{figure*}%[!ht]
  \centering
  \includegraphics[width=380px]{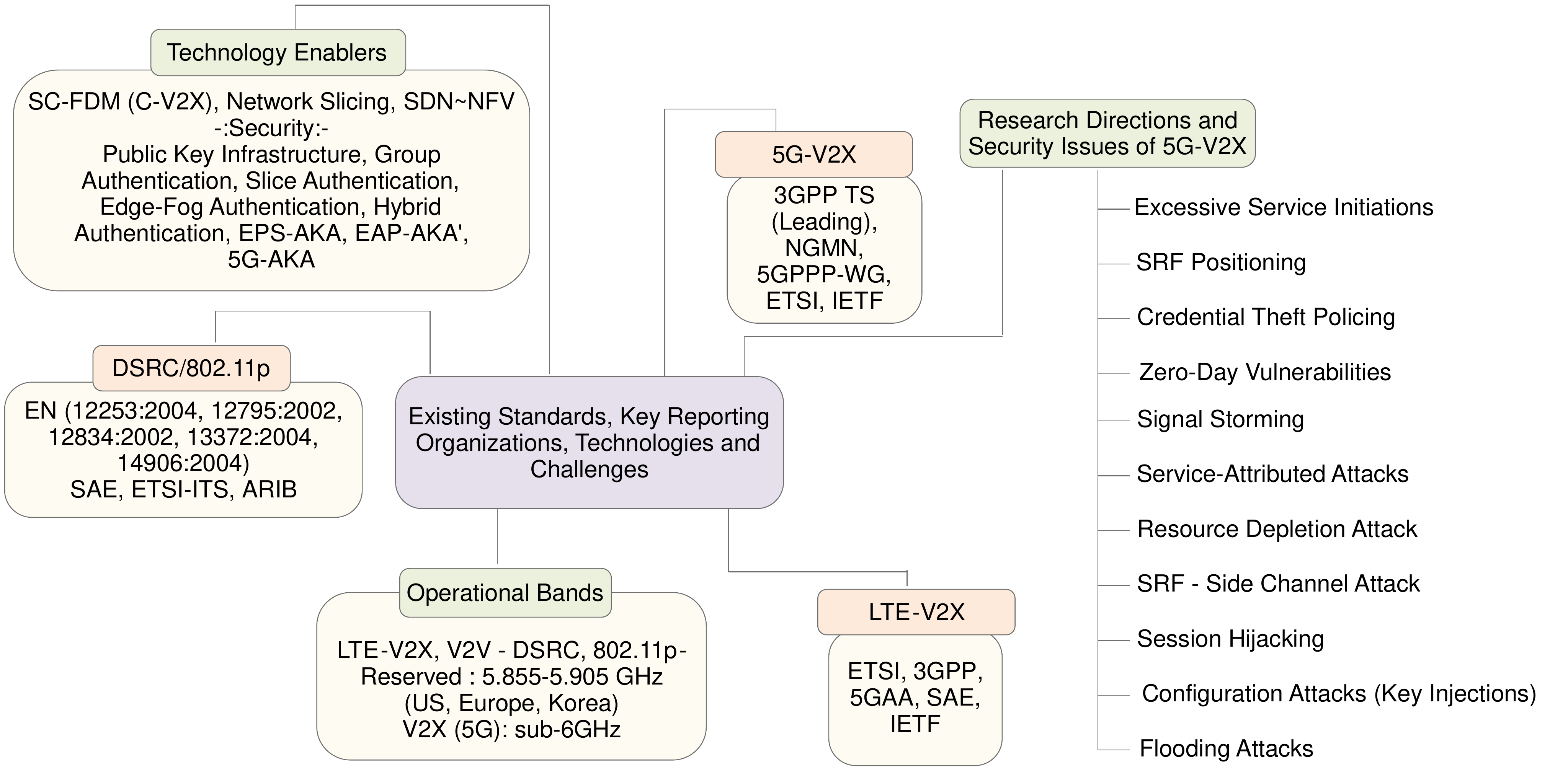}
  \caption{An illustration of existing standards, key reporting organizations, technologies and research challenges particularly emphasizing the proposed SRF-based 5G-V2X.}\label{fig5}
\end{figure*}

\subsection{5G-V2X Authentication}
This conceptualized architecture can be used to support the host as well as network-initiated authentications. For V2X security, the proposed architecture uses $K_{X}$ and $K_{V}$ for edge-initiated authentication. Both these keys are derived at the edge and it prevents any long-distance transmissions, which also helps to attain optimized routes for the derivation of keys as well as V2X authentication. The location and trajectory initiated key generation also reduces the signaling overheads as $K_{SRF}$ alone can be used to generate the edge-side keys.

SRF-based architecture is secure for most of the issues that are presented in Fig.~\ref{fig3}. For the authentication and authorization, SRF derives its keys from AMF, which follows session-wise key generation, thus, maintaining the freshness throughout the connectivity. Even if the network synchronization is disturbed for the vehicles, the SRF maintains an independent connection to the core security functions, allowing zero-drop during re-verification. Most dominantly, SRF helps to overcome issues related to certificate forgery and also allows expensive public-key operations to be used by providing a short-pass between the vehicles and everything. The vehicles can use dual authentication through SRF, which allows partial authentication with the gNB or nearby sensor and partial authentication with the core functions, once the vehicle-services are initialized.

In the intra-mode, when the vehicles operate in the periphery of gNB, SV, or a sensor, no additional mechanisms or key exchanges are required for re-authentication as $K_{V}$ is securely derived from $K_{SRF}$. The rapid changes to the network are also handled by the re-authentication with the SRF, which also prevents issues related to access management. For the inter-mode operations (handover), AMF accounts for the security of SRFs and prevents re-authentication by relying on SRF to check the validity of the vehicle under movement.

All these operations help to decrease the total cost of operations, which is measured on the basis of the number of hops to be traversed for generating keys, especially during the re-authentication. Moreover, the end to end security as well as the backward security can easily be observed by including the SRF in the existing architecture. SRF also exhibits the control properties which can be extended through a different framework for attaining UD-US mobility management. Furthermore, SRF, through the disintegration of authentication and mobility management, allows extensive privacy and anonymous operations along with highly secure network management. The disintegrated operations also reduce the network stress in terms of overheads by dividing the user-side management functionaries.

%The proposed architecture can considerably reduce the burden on the core for authenticating vehicles during mobility.

To present this, standard EAP-AKA' is used for authentication when a vehicle moves across the terminals in the traditional~\cite{3GPPTS} and the proposed setup. The proposed architecture brings authentication near to the vehicle at the edge leading to conservation of 2.5\% to 11.3\% signaling overheads, as shown in Fig.~\ref{fig6}, by utilizing the computational model given in~\cite{sharma2018secure}. The signaling overheads are reduced based on the intermediate hops involved in authentication. The output for EAP-AKA' shows lesser improvement as the protocol is driven by core security functions. However, to fully utilize the proposed architecture, it is recommended to consider developing novel protocols that can enhance the performance to a large extent at a similar strength of security.

\section{Discussions and Research Challenges}
Security concerns of C-V2X are dominated by the type of architecture used for deploying devices and entities up to the core. Especially for 5G-V2X, there are some additional security concerns, such as service-based accessibility, signal storming, and edge-based authentications. These issues did not prevail much in LTE-V2X; however, for NS-5G-V2X, these become dominant performance affecting concerns for LTE equipment. Further details can be seen in Fig.~\ref{fig5}.

%Irrespective of the mode of deployment, it becomes inevitably important to curb their presence and prevent the network from uncertain failures and threats. The majority of the research aspects are similar to those discussed in earlier parts of this article, and this section mainly discusses the concerns and challenges related to conceptualized architecture.

Some of the key research challenges and open issues, which can be targeted as a part of future work, while using the conceptualized architecture, are presented below:
 \begin{itemize}
   \item Prevention of excessive service initiations: When an attacker is able to violate the SRF, it may initiate multiple services to interrupt the V2X operations.
   \item SRF positioning: The positioning of the SRF function is an optimization issue and it may vary from one scenario to the next. In some cases, where OEM wants direct control of the vehicles, SRF functions need to be placed near the OEM, which will violate the principles of edge-computing. Thus, its positioning and selection of key-relaying solutions are major concerns to be resolved.
   \item Accurate sensor relaying: The location and trajectory-based key generations are based on accurate sensor readings. Thus, it is desired that vehicles' data is accurately retrieved under all circumstances.
   \item Credential theft: In case of false requests from the vehicles, the exposure of $K_{V}$ may pose threats to $K_{X}$, which is the key of the devices with which the vehicles communicate directly. Thus, this will involve re-authentication, but identification of an instance of re-authentication is tedious, and it is desired to develop strategies for credential theft. Solutions like self-evaluations and introduction of self-checking logic can help to facilitate these requirements whenever the vehicles are initiated in the network.
   \item Configuration attacks: These attacks are most dominant for NS-5G-V2X, as V2V/P broadcasts can be used to mislead the receiving entity to make wrong decisions. These attacks also pave a way for routing attacks as well as session hijacking.
   \item Perfect forward secrecy: This issue is applicable to the majority of the networks as no concurrent approach can provide perfect forward secrecy at an efficient rate. However, with the use of certain technologies and protocols, it can be achieved on the backbone of the conceptualized architecture by utilizing several instances of SRF.
   \item Insider threats and zero-day attacks: Privacy and anonymity are affected most by insider threats and potential zero-day vulnerabilities. Both these tend to expose the entire network and share the key-exchange phenomenon to outer entities, which can launch attacks to misled the vehicles~\cite{zhang2017security}. Thus, managing insider threats and developing strategies to prevent zero-day attacks, by understanding the window of vulnerability especially for V2X, are major challenges to resolve in 5G-V2X.
 \end{itemize}
\section{Conclusion}
This article presents an overview of C-V2X technologies and standards while focusing on the current situations of LTE-V2X and 5G-V2X. Several use-cases, service supports, and security requirements are discussed in detail. Issues related to existing 5G-V2X based on standalone as well as non-standalone are presented through comparisons. A conceptualized Security Reflex Function (SRF)-based architecture is also presented, which aims to reduce the burden of secure mobility management of vehicles in 5G-V2X. In addition, various open issues and research directions are discussed which help to understand the current aspects of 5G-V2X and its security alongside the usability of the conceptualized architecture.

\bibliographystyle{ieeetr}
\bibliography{v2x_related}
%\bibliographystyle{ieeetran}
%\bibliography{bibfile}
\end{document}